**Localization of Yttrium Segregation within YSZ Grain Boundary Dislocation Cores**

*Gabriel Sánchez-Santolino\*, Juan Salafranca, Sokrates T. Pantelides, Stephen J. Pennycook, Carlos León and María Varela*

Dr. G. Sánchez-Santolino, Dr. J. Salafranca, Prof. C. León, Prof. M. Varela
GFMC, Dept. de Física de Materiales, Universidad Complutense de Madrid
28040 Madrid, Spain.
E-mail: gabriel.sanchez.sant@ucm.es

Dr. G. Sánchez-Santolino, Prof. M. Varela
Instituto Pluridisciplinar, Universidad Complutense de Madrid,
28040 Madrid, Spain.

Dr. G. Sánchez-Santolino
Present address: 2D-Foundry Group, Instituto de Ciencia de Materiales de Madrid ICMM-CSIC, 28049 Madrid, Spain.

Prof. S. T. Pantelides
Department of Physics and Astronomy, Vanderbilt University,
Nashville, TN 37235, USA

Porf. S. J. Pennycook
Department of Materials Science & Engineering, National University of Singapore, Singapore 117575.

Prof. C. León
Unidad Asociada UCM/CSIC, "Laboratorio de Heteroestructuras con Aplicación en Spintrónica".
GFMC, Instituto de Magnetismo Aplicado, Universidad Complutense de Madrid,
28040 Madrid, Spain.

**Abstract**

Ionic conductivity blocking at grain boundaries in polycrystalline electrolytes is one of the main obstacles that need to be overcome in order to improve the performance of solid state fuel cells and batteries. To this aim, harnessing the physical properties of grain boundaries in ionic conducting materials such as yttria stabilized zirconia (YSZ) down to the atomic scale arises as a greatly important task. Here we present a structural and compositional analysis of a single grain boundary in a 9 mol% yttria content YSZ bicrystal by means of aberration corrected scanning transmission electron microscopy. Our studies combine strain and compositional atomic resolution analysis with density-functional-theory calculations in order to find a preferential segregation of yttrium to the expansive atomic sites at the grain boundary dislocation cores. These results address a crucial step towards the understanding of the physical properties of grain boundaries down to atomic dimensions.

**1. Introduction**

The functionality of materials and devices is critically determined by the physical properties of small active regions such as surfaces, interfaces and grain boundaries. Chemical inhomogeneities within these regions, such as those produced by segregation of dopants to grain boundaries determine the physical properties of the boundary and thus the macroscopic response of materials and devices[1,2]. In particular, the performance of ionic conducting materials is greatly affected by the presence of such interfaces[3–5]. Polycrystalline ionic conductors are commonly used as electrolytes of solid oxide fuel cells (SOFCs), so ionic blocking processes at grain boundaries constitute one of the main obstacles demining their performance[6,7]. Therefore, it is crucial to understand the mechanisms controlling the segregation of dopants to grain boundaries as well as the ensuing structural deformations[8,9]. This is a particularly pressing need when studying oxygen ionic conductors such as yttria $Y_2O_3$ stabilized zirconia $ZrO_2$ (YSZ). YSZ is a prototypical ionic conductor used as electrolyte in SOFCs[10,11]. Zirconia can be stabilized into a fluorite cubic structure at high temperatures, when vacancies at the oxygen anion sites tend to form favoring the oxygen ionic conductivity. When doping with yttria, the lower oxidation state and larger radius of the $Y^{3+}$ ions promote the introduction of oxygen vacancies into the crystal and stabilize the high temperature cubic fluorite structure at room temperature. These characteristics make of YSZ a good candidate as an oxygen ionic conductor. But strain fields and structural deviations near extended defects such as grain boundaries have complex effects on the local composition and, thus, on the carrier doping and mobility.

In this scenario, the determination of preferential atomic sites for dopant segregation at grain boundaries with atomic precision is still challenging. Real space probes such as aberration corrected scanning transmission electron microscopy (STEM) combined with electron energy-loss spectroscopy (EELS) has proven to be a much powerful technique to study materials even down to the single atom level[12–15]. Thus, the capabilities of STEM-EELS for determining the atomic structure and composition of grain boundaries in an atomic-site-by-atomic-site fashion provide a unique tool in order to address these issues. In this work we report an atomic resolution strain analysis of grain boundaries in ionic conductors via atomic resolution STEM-EELS and relate our findings to the chemical variations taking place. We have studied an yttria 9 mol% YSZ bicrystal with a single grain boundary. Previously, we found a significant yttrium segregation to the grain boundary dislocation cores[16], although the specific sites of segregation or any dependence on the local strain were not determined. Here we report an inhomogeneous strain landscape around the boundary in the form of compressive and expansive sites at the dislocation cores along the grain boundary by detailed measurements of the local strains. Then, we use atomic resolution EELS to probe the chemical variations at the grain boundary, identifying the dependence of yttrium segregation with the local strain and explain the local properties of the system by density functional theory calculations.

## 2. Experimental Section

The YSZ bicrystal was made by means of solid phase intergrowth. For this aim, two single crystals are cut to obtain the surface in the desired orientations, and then, aligned together and annealed under pressure in an ultra-high vacuum environment. Usual annealing conditions are 1873K for 15h. The sample studied is a commercial 9 mol% yttria YSZ bi-crystal with a symmetrical 33° [001] tilt grain boundary acquired from MaTeck GmbH. Suitable specimens for STEM-EELS were prepared by conventional methods: grinding and ion milling. Samples were coated with a 1 nm thick layer of Ir to prevent charging. The measurements were carried out in a Nion UltraSTEM200 operated at 200 kV and equipped with a fifth order aberration corrector. For strain analysis, the Peak Pairs Analysis (PPA) software package (HREM Research) for Digital Micrograph was used[17]. This approach has successfully served to quantify local strain distortions in atomic resolution STEM images[18]. For the compositional analysis, a series of EEL spectrum images were acquired using an Enfinium Gatan spectrometer, which allows simultaneous acquisition of low (O $K$ edge) and high (Zr and Y $L_{2,3}$ edges) energy-loss edges. For spectrum imaging, the electron probe is raster scanned along the region of interest and an EEL spectrum is acquired in every pixel. The EEL spectrum images were acquired with a 1 eV/channel dispersion and an exposure time of 0.05 seconds/pixel. Principal component analysis was used to remove random noise[19]. Density functional theory calculations were used to obtain the theoretical energies and atomic structure of the bicrystal. For the calculations, the projector augmented-wave method (PAW)[20,21] as implemented in the VASP code[22,23] was used.

## 3. Results and Discussion

### 3.1. YSZ bicrystal grain boundary characterization

The orientation relationship between the two single crystals and the structural quality of the sample was determined from atomic resolution STEM images. **Figure 1** shows a high angle annular dark field (HAADF) image of the grain boundary region where a regular array of evenly distributed dislocation cores can be observed. Neither disordered nor amorphous structures are present, indicating that the boundary is perfectly joined at the atomic level. Due to the highly insulating nature of the sample and to prevent charging effects, the specimen was coated with a thin iridium layer, which appears as areas with a brighter, inhomogeneous contrast in the HAADF image. Figure 1(b) shows the fast Fourier transform (FFT) of the image in (a). The high symmetry of the pattern indicates the perfect alignment between the two single crystals. The crystallographic orientation relationship between the two sides of the bicrystal grain boundary was measured from the FFT. We obtain a 68° angle between the symmetric reflections of both crystals, which is close to the expected value.

### 3.2. Atomic strain analysis at the grain boundary dislocation cores

We use now the PPA routines[17] to determine the local strains at the grain boundary region. The presence of an array of dislocation cores along the grain boundary is known to produce a highly inhomogeneous strain field within the neighboring area, with the formation of compressive and expansive strain regions[24]. Usually, larger ions occupy the expansive strain sites and smaller ones the compressive ones in order to help release any local strains[25], so in our case, yttrium is expected to occupy the expansive sites when segregating to the grain boundary. **Figure 2**(a) shows an atomic resolution HAADF image used for the strain analysis. As the reference lattice is not equal for the whole image due to the relative miss-orientation between the two sides of the bicrystal, we choose two symmetrical directions between both sides as the principal vectors for the strain analysis. The [100] directions for each side of grain boundary, used as the principal vectors for the PPA analysis, are marked with yellow arrows in the image. As reference for our calculation we have chosen an extensive region containing both sides of the grain boundary. The resultant strain map describes the difference in percentage of the mean nearest neighbor's bond lengths along the chosen principal vectors of the atomic sites in our region of interest in comparison with the selected reference region. We find a strong expansive strain localized at the dislocation cores, as depicted by the modulus of the strain tensor $d_{xy} = \sqrt{\varepsilon_{xx} + \varepsilon_y}$ shown in Figure 2(b). We have superimposed the strain tensor map to the HAADF image in order to easily determine the expansive and compressive sites at the grain boundary dislocation cores, which appear marked with white and black dots in the figure, respectively. This way we identify three expansive atomic sites and one compressive site per core, as explained in the schematic on Figure 2(d). We measure the mean dilatation ($d_{xy}$) at each of the expansive and compressive sites along all the dislocation cores within the analyzed region obtaining the following averaged values: $E_1$= +15.3%, $E_2$= +10.7%, $E_3$= +23.8% and $C_1$= -2.7%. It is important to note that the current results reflect a two-dimensional measurement of a three-dimensional structure and the possible strain distortions along the beam direction are not considered. Nonetheless recent advances in three-dimensional atomic imaging of crystalline structures by techniques such as electron tomography[26,27], in-line holography[28] and large-angle illumination STEM[29] bring closer the much interesting possibility of experimentally determining the three-dimensional structure of grain boundary dislocation cores with atomic precision.

**3.3. Mapping the chemical variation at the YSZ grain boundary**

In order to correlate the structural analysis of the grain boundary with any chemical variations taking place at the dislocation cores we use next atomic resolution electron energy-loss spectroscopy techniques. EELS techniques allow identifying compositional changes for the specific atomic sites within the grain boundary. This is not an easy task because the signal arising from inelastic scattering, as the one used for EEL spectroscopy, is broadened by the spreading of the probe as it passes through a material (dechanneling) and by the delocalization of the ionization events producing the signal itself[30,31]. Thus, the signal acquired when the probe is placed over a specific atomic site also contains information originating from the neighboring columns. Due to the lower delocalization of the $L_{2,3}$ ionization interaction, we choose to analyze the Y and Zr $L_{2,3}$ edges at 2080 eV and 2200 eV respectively, which will allow obtaining more accurate results. An EEL spectrum image was acquired from the area marked with a yellow box in **Figure 3**(a). The EELS integrated intensity maps for all atomic species are shown in Figure 3(b). Noticeable compositional changes are observed with the periodicity of the grain boundary dislocation cores[32,33]. We find a higher Y signal at the dislocation cores along with a reduction of both Zr and O intensities. Such changes in the EELS signal are highly localized on the grain boundary region.

In order to quantify any chemical changes at the grain boundary a quantification of the relative composition was carried out. It is important to note that the structural distortions at the grain boundary affect the signal acquired by the EELS detectors when compared to the bulk. Grain boundaries present a lower atomic density due to partially occupied columns, as hinted by the reduced contrast of some atomic columns at the dislocation cores in the HAADF images. Also, grain boundaries have a disordered structure which will affect the way the electron beam channels through the sample[34]. Thus, it is often preferable in general to perform the quantitative composition analysis in reduced channeling conditions, which is accomplished by slightly tilting the crystal out of axis. Indeed in previous works based on this approach we have found a difference of a few percent between on axis and out of axis conditions[16]. However, in our case we aim to compare the strain analysis above with the composition of specific sites within the grain boundary, so it is necessary to acquire the EEL spectrum images on-axis conditions. It should be then noted that our quantitative values can be slightly affected by the not optimal conditions but the overall trends found should be robust. That said, we employed the usual EELS quantification routines available in the Gatan microscope suite Digital Micrograph in order to obtain the relative compositional maps between the three atomic species in the sample. The chemical quantification is then calculated individually for every pixel in a pixel-by-pixel basis and the resultant values are the relative concentration values for each pixel of the spectrum image and do not depend on neighboring values. Thus, the slight sample drift observed does not affect the relative concentration results presented in our manuscript. The resultant relative maps for Y, Zr and O, shown in quantitative color scales are shown at the bottom row. They exhibit how the value of the Y atomic concentration, which for a bulk 9 mol% yttria ($Y_2O_3$) stabilized zirconia corresponds to a 5.7% yttrium relative content to the total O, Zr and Y concentrations, almost doubles on the GB dislocation cores. This result is expected as Y is known to segregate to the grain boundary region[34–37]. The experimentally measured zirconium and oxygen relative compositions both decrease at the grain boundary dislocation cores, but not as much as hinted by the direct EELS intensity maps. The oxygen relative composition decrease can be related to the accumulation of oxygen vacancies. We could not detect any variations associated with nanometric wide oxygen vacancy depleted space charge regions[38,39] in agreement with a previous report in this YSZ grain boundary[16]. This is still a subject of controversy and several theoretical studies have explained the chemistry of YSZ grain boundaries without the formation of any layer depleted of oxygen vacancies and emphasize the importance of resolving the specific atomic arrangements[40,41].

## 3.4. Determination of the Y segregation with atomic site precision

The quantitative results obtained from the atomic resolution EEL spectrum images allow correlating the compositional changes taking place on the dislocation cores with the structural distortions found in the strain analysis. We will focus on the expansive and compressive sites of the grain boundary, as determined by the strain analysis. **Figure 4**(a) shows the atomic resolution annular dark field image acquired simultaneously with the spectrum image shown in Figure 3. If we plot an overlay including both the HAADF and the Y integrated intensity map together (Figure 4(b)) it is possible to observe how the Y content increase is highly localized on a single atomic column within the whole dislocation core structure, pointing to a preferential segregation site. On the right panel we show a graph with the quantitative relative compositional values for Y, Zr and O at the expansive and compressive sites in comparison with the bulk values, marked with horizontal lines (green for O, blue for Zr, red for Y). The values at each site are obtained by averaging over the four dislocation cores probed in the spectrum image region. These results exhibit a clear difference between sites. At the compressive site $C_1$ and expansive sites $E_1$ and $E_2$ (the ones under less expansive strain) we measure relatively small concentration changes with values close to the bulk for all Y, Zr and O and in all cases, within our experimental error. On the other hand, we find much stronger variations at the expansive site $E_3$. The yttrium content in this atomic site doubles the one in the bulk, suggesting a highly preferential segregation to the position under higher expansive strain. In fact, density-functional theory calculations indicate that Y segregates to the grain boundary with a segregation energy of 2.9 eV[16]. Y preferentially substitutes Zr at the $E_3$ expansive site of the grain boundary, in agreement with experiments, and despite the larger coordination number. Such results were already reported in our previous publication, and here they are only used to discuss new experimental evidence in the context of those previous calculations. This result is consistent with the decrease in oxygen concentration measured at this site. $Y^{3+}$ ions lower the energy formation for oxygen vacancies[16] and hence, promote the localization of vacancies at the grain boundary, keeping charge neutrality and releasing strain at the core[42].

**4. Conclusions and Summary**

In summary, we have analyzed the structural distortions and chemical variations taking place at a 9 mol% yttria content YSZ bi-crystal with a symmetrical 33° [001] tilt grain boundary. Using peak pairs analysis we determine the expansive and compressive sites at the grain boundary dislocation cores with atomic column precision. By means of electron energy-loss spectroscopy we map the composition at the grain boundary region. We find variations with the periodicity of the dislocation cores where Y strongly increases in composition at the boundary. Comparing the annular dark field image with the relative concentration Y map we identify the segregation of yttrium to a single atomic column. This preferential segregation site, which corresponds to the atomic position under the higher expansive strain is confirmed by density functional theory calculations. The highly localized Y segregation explains the confinement of oxygen vacancies at the dislocation cores and is identified as the main mechanism underlying strain relaxation near the grain boundary, which should be taken into account when explaining the properties of these systems.

**Acknowledgements**
We acknowledge financial support by the MINECO-FEDER MAT2015-66888-C3-3-R and the ERC PoC2016 POLAR-EM grant. G.S-S was supported by the Juan de la Cierva fellowship FJCI-2015-25427 (MINECO-Spain). The authors thank Dr. Hugo Schlich from Mateck GmbH for providing information about the process of bicrystals production, Masashi

**References**

1. R.F. Klie, J.P. Buban, M. Varela, A. Franceschetti, C. Jooss, Y. Zhu, N.D. Browning, S.T. Pantelides, S.J. Pennycook. Nature **2005**, 435, 475.

2. J.P. Buban, K. Matsunaga, J. Chen, N. Shibata, W.Y. Ching, T. Yamamoto, Y. Ikuhara. Science **2006**, 311, 212.

3. X. Guo. Solid State Ionics 1995, 81, 235.

4. D. Owen, A. Chokshi. Acta Mater. **1998**, 46, 667.

5. M. Aoki, Y.-M. Chiang, I. Kosacki, L.J.-R. Lee, H. Tuller, Y. Liu. J. Am. Ceram. Soc. **1996**, 79, 1169.

6. R.M. Ormerod. Chem. Soc. Rev. **2003**, 32, 17.

7. J.B. Goodenough. Annu. Rev. Mater. Res. **2003**, 33, 91.

8. E. Dickey, X. Fan, S. Pennycook. J. Am. Ceram. Soc. **2001**, 68.

9. J.-Y. Roh, Y. Sato, Y. Ikuhara. Dickey B, ed. J. Am. Ceram. Soc. **2015**, 98, 1932.

10. B.C.H. Steele, A. Heinzel. Nature **2001**, 414, 345.

11. J.W. Fergus. J. Power Sources **2006**, 162, 30.

12. K. Suenaga, M. Tencé, C. Mory, C. Colliex, H. Kato, T. Okazaki, H. Shinohara, K. Hirahara, S. Bandow, S. Iijima. Science **2000**, 290, 2280.

13. M. Varela, S.D. Findlay, A.R. Lupini, H.M. Christen, A.Y. Borisevich, N. Dellby, O.L. Krivanek, P.D. Nellist, M.P. Oxley, L.J. Allen, S.J. Pennycook. Phys. Rev. Lett. **2004**, 92, 95502.

14. O.L. Krivanek, M.F. Chisholm, V. Nicolosi, T.J. Pennycook, G.J. Corbin, N. Dellby, M.F. Murfitt, C.S. Own, Z.S. Szilagyi, M.P. Oxley, S.T. Pantelides, S.J. Pennycook. Nature **2010**, 464, 571.

15. N. Shibata, T. Seki, G. Sánchez-Santolino, S.D. Findlay, Y. Kohno, T. Matsumoto, R. Ishikawa, Y. Ikuhara. Nat. Commun. **2017**, 8, 15631.

16. M.A. Frechero, M. Rocci, G. Sánchez-Santolino, A. Kumar, J. Salafranca, R. Schmidt, M.R. Díaz-Guillén, O.J. Durá, A. Rivera-Calzada, R. Mishra, S. Jesse, S.T. Pantelides, S. V. Kalinin, M. Varela, S.J. Pennycook, J. Santamaria, C. Leon. Sci. Rep. **2015**, 5, 17229.

17. P.L. Galindo, S. Kret, A.M. Sanchez, J.-Y. Laval, A. Yáñez, J. Pizarro, E. Guerrero, T. Ben, S.I. Molina. Ultramicroscopy **2007**, 107, 1186.

18. A. Llordés, A. Palau, J. Gázquez, M. Coll, R. Vlad, A. Pomar, J. Arbiol, R. Guzmán, S. Ye, V. Rouco, F. Sandiumenge, S. Ricart, T. Puig, M. Varela, D. Chateigner, J. Vanacken, J. Gutiérrez, V. Moshchalkov, G. Deutscher, C. Magen, X. Obradors. Nat. Mater. **2012**, 11, 329.

19. M. Bosman, M. Watanabe, D.T.L. Alexander, V.J. Keast. Ultramicroscopy **2006**, 106, 1024.

20. P.E. Blöchl. Phys. Rev. B **1994**, 50, 17953.

21. G. Kresse. Phys. Rev. B **1999**, 59, 1758.

22. G. Kresse, J. Hafner. Phys. Rev. B **1993**, 47, 558.

23. G. Kresse. Phys. Rev. B **1996**, 54, 11169.

24. M.J. Hÿtch, J.-L. Putaux, J. Thibault. Philos. Mag. **2006**, 86, 4641.

25. J. Takahashi, K. Kawakami, J. Hamada, K. Kimura. Acta Mater. **2016**, 107, 415.

26. S. Van Aert, K. J. Batenburg, M. D. Rossell, R. Erni, G. Van Tendeloo. Nature **2011**, 470, 374.

27. C.-C. Chen, C. Zhu, E. R. White, C.-Y. Chiu, M. C. Scott, B. C. Regan, L. D. Marks, Y. Huang, J. Miao. Nature **2013**, 496, 74.

28. F.-R. Chen, D. Van Dyck, C. Kisielowski.Nat. Commun. **2016**, 7, 10603.

29. R. Ishikawa, A. R. Lupini, Y. Hinuma, S. J. Pennycook. Ultramicroscopy **2015**, 151, 122.

30. L. Allen, S. Findlay, A. Lupini, M. Oxley, S. Pennycook. Phys. Rev. Lett. **2003**, 91, 1.

31. M. Zacate. Solid State Ionics **2000**, 128, 243.

32. Y. Yan, M. Chisholm, G. Duscher, A. Maiti, S. Pennycook, S. Pantelides. Phys. Rev. Lett. **1998**, 81, 3675.

33. Z. Wang, M. Saito, K.P. McKenna, L. Gu, S. Tsukimoto, A.L. Shluger, Y. Ikuhara. Nature **2011**, 479, 380.

34. B. Feng, N.R. Lugg, A. Kumamoto, Y. Ikuhara, N. Shibata. ACS Nano **2017**, 11, 11376.

35. Y. Lei, Y. Ito, N.D. Browning, T.J. Mazanec. J. Am. Ceram. Soc. **2002**, 85, 2359.

36. K. Matsui, H. Yoshida, Y. Ikuhara. Acta Mater. **2008**, 56, 1315.

37. B. Feng, T. Yokoi, A. Kumamoto, M. Yoshiya, Y. Ikuhara, N. Shibata. Nat. Commun. **2016**, 7, 11079.

38. X. Guo, R. Waser. Prog. Mater. Sci. **2006**, 51, 151.

39. R.L. González-Romero, J.J. Meléndez, D. Gómez-García, F.L. Cumbrera, A. Domínguez-Rodríguez. Solid State Ionics **2013**, 237, 8.

40. M. Yoshiya, T. Oyama. J. Mater. Sci. **2011**, 26, 4176

41. T. Yokoi, M. Yoshiya, H. Yasuda, Langmuir **2014**, 30,47, 14179.

42. M.O. Zacate, L. Minervini, D.J. Bradfield, R.W. Grimes, K.E. Sickafus. Solid State Ionics **2000**, 128, 243.

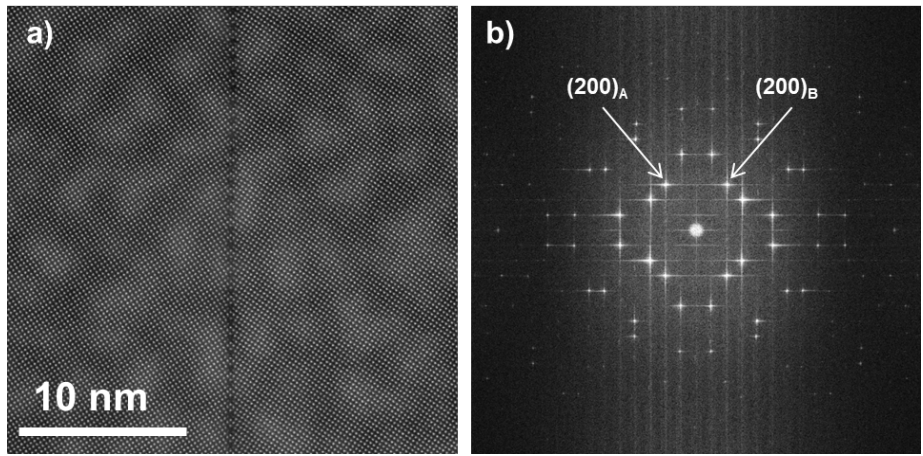

**Figure 1.** (a) High angle annular dark field image of a YSZ bicrystal with a 33° symmetric tilt [001] grain boundary. (b) Fast Fourier transform of the image in (a) where two symmetric reflections are marked with white arrows.

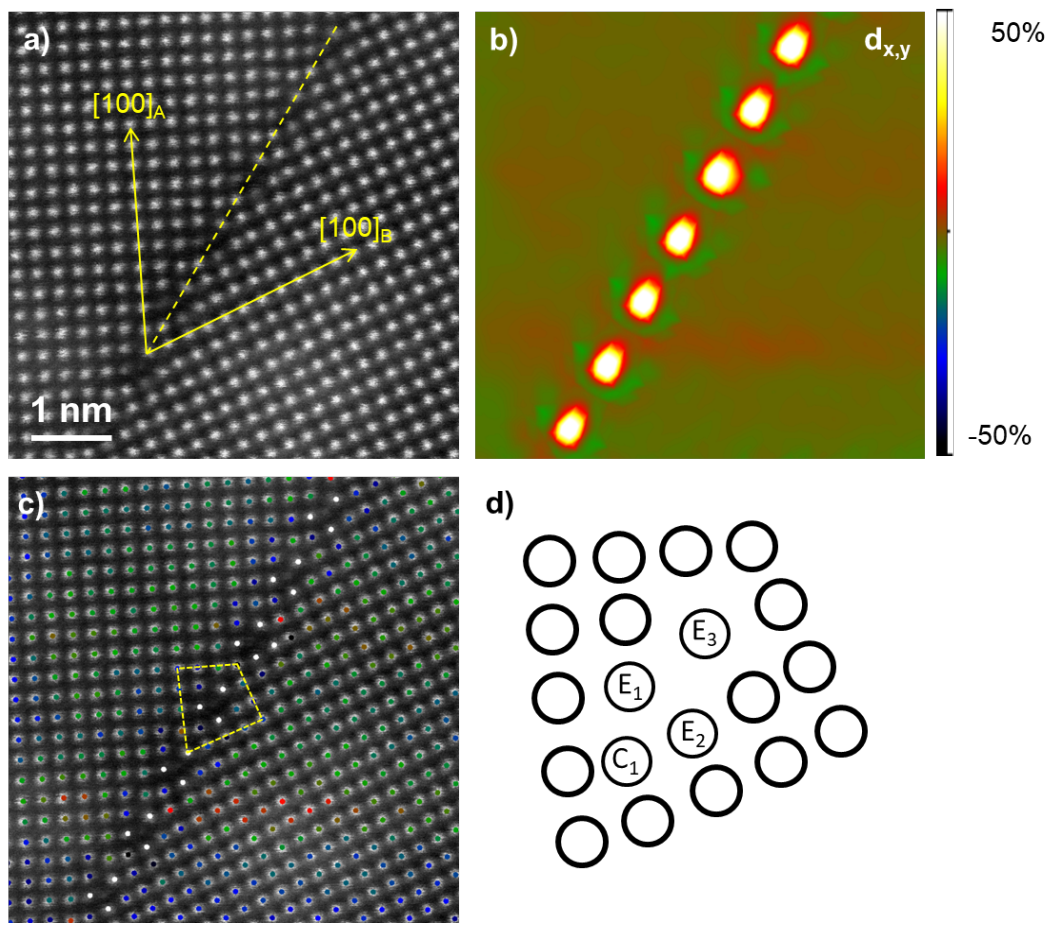

**Figure 2.** (a) HAADF image of the grain boundary region. The two symmetric directions used as the principal vectors for the PPA analysis are marked with yellow arrows on the image. (b) Mean dilatation ($d_{xy}$) map of the grain boundary area in %100 obtained from PPA[17] analysis. (c) $d_{xy}$ shown as an overlay to the HAADF image in the same color scale, where the expansive and compressive sites of the grain boundary dislocation cores are marked with red and blue arrows respectively. (d) Schematic of the dislocation cores, as marked with dashed yellow lines in (c), indicating the three expansive sites $E_1$, $E_2$, $E_3$ and the compressive site $C_1$ as found by PPA analysis.

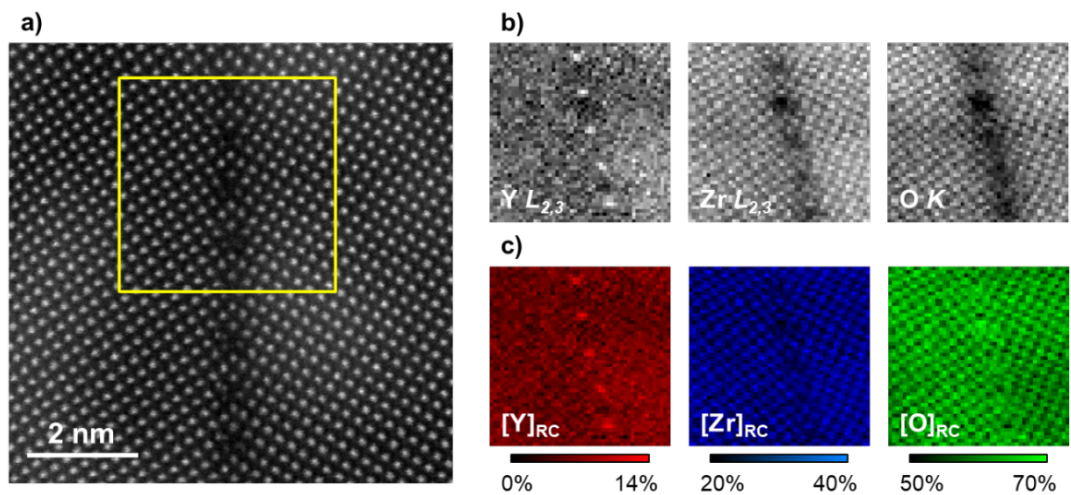

**Figure 3.** (a) HAADF image of the grain boundary region. A yellow box depicts the area where an EEL spectrum image was acquired. (b) Y $L_{2,3}$, Zr $L_{2,3}$ and O $K$ edges integrated intensity maps produced by integrating the intensity under the edges after removing the background using a power law. (c) Relative composition maps for Y, Zr and O depicting the compositional changes taken place at the grain boundary region. A slight sample drift during acquisition can be observed.

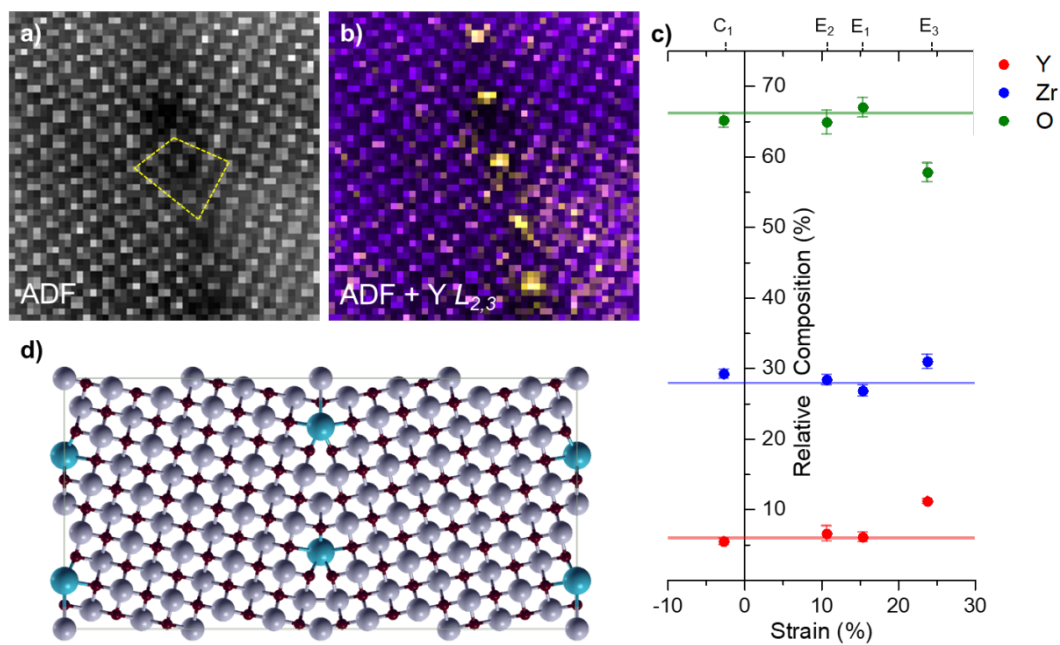

**Figure 4.** (a) ADF image acquired simultaneously with the EEL spectrum image in Figure 3. The dislocation core structure, as in Figure 2, is marked with dashed yellow lines. (b) Combined simultaneous ADF (purple) and Y $L_{2,3}$ (yellow) map in which the Y preferential segregation to the expansive site at the dislocation core can be easily observed. (c) Graph showing the results of the relative composition values for Y, Zr and O at the expansive and compressive atomic sites of the grain boundary dislocation cores as function of the measured strain. The horizontal lines indicate the bulk values. The error bars indicate the standard deviation of the set of values measured on the dislocations cores present in the spectrum image region. (d) Super cell of an $Y_2O_3$ stabilized $ZrO_2$ grain boundary used in the theoretical calculations. Zr is represented by large grey spheres, O by small dark-red spheres and substitutional Y by blue spheres.